\documentclass{aa}  
\usepackage{graphicx}
\usepackage{xcolor}
\usepackage{mathrsfs}
\usepackage{txfonts}
\begin{document} 

\def\percc{\rm cm^{-3}}
\def\kms{\rm km\,s^{-1}}
\def\persqcm{\rm cm^{-2}}
\def\micron{\mu{\rm m}}
\def\htwo{\rm H_2}
\def\diaz{\rm N_2H^+}
\def\ddiaz{\rm N_2D^+}
\def\htwodplus{\rm H_2D^+}
\def\dtwohplus{\rm D_2H^+}

   \title{Hunting pre-stellar cores with APEX\\ IRAS16293E (Oph464)}

\author{S. Spezzano\inst{1}  \and E. Redaelli\inst{1,2} \and P. Caselli\inst{1} \and O. Sipil\"a\inst{1} \and J. Harju\inst{1},\inst{3}\and F. Lique\inst{4} \and D. Arzoumanian\inst{5} \and J.~E. Pineda\inst{1} \and F. Wyrowski\inst{6} \and A. Belloche\inst{6}}  

\institute{Max-Planck-Institut f\"ur Extraterrestrische Physik, Giessenbachstrasse 1, 85748 Garching, Germany \and 
European Southern Observatory, Karl-Schwarzschild-Strasse 2, 85748 Garching, Germany \and 
Department of Physics, P.O. Box 64, FI-00014, University of Helsinki, Finland \and Univ. Rennes, CNRS, IPR (Institut de Physique de Rennes) – UMR 6251, 35000 Rennes, France \and National Astronomical Observatory of Japan, Osawa 2-21-1, Mitaka, Tokyo 181-8588, Japan \and Max-Planck-Institut f\"ur Radioastronomie, Auf dem H\"ugel 69, 53121 Bonn, Germany}

\abstract 
{Pre-stellar cores are the first steps in the process of star and planet formation. However, the dynamical and chemical evolution of pre-stellar cores is still not well understood. Our partial knowledge on the chemical and physical structure of pre-stellar cores, as well as how they are fed and influenced by the surrounding environment, limits the level of knowledge that we can achieve on later stages in the star and planet formation process, from protostellar cores to exoplanets.}
{We aim at estimating the central density of the pre-stellar core IRAS16293E ($Herschel$ Gould Belt Survey core Oph464) and at carrying out an inventory of molecular species towards the density peak of the core.}
{We observed high-$J$ rotational transitions of N$_2$H$^+$ and N$_2$D$^+$, and several other molecular lines towards the dust emission peak using the Atacama Pathfinder EXperiment (APEX) telescope, and derived the density and temperature profiles of the core using far-infrared surface brightness maps from $Herschel$. The N$_2$H$^+$ and N$_2$D$^+$ lines were analysed by non-LTE radiative transfer modelling.}
{Our best-fit core model consists in a static inner region, embedded in an infalling
envelope with an inner radius of approximately 3000 au (21" at 141 pc). The observed high-J lines of N$_2$H$^+$ and N$_2$D$^+$ (with critical densities greater than 10$^6$ cm$^{-3}$) turn out to be very sensitive to depletion; the present single-dish observations are best explained with no depletion of N$_2$H$^+$ and N$_2$D$^+$ in the inner core. The N$_2$D$^+$/N$_2$H$^+$ ratio that best reproduces our observations is 0.44, one of the largest observed to date in pre-stellar cores. Additionally, half of the molecules that we observed are deuterated isotopologues, confirming the high-level of deuteration towards this source.} 
{Non-LTE radiative transfer modelling of N$_2$H$^+$ and N$_2$D$^+$ lines proved to be an excellent diagnostic of the chemical structure (i.e. molecular freeze-out) and dynamics (infall velocity profile) of a pre-stellar core. Probing the physical conditions immediately before the protostellar collapse is a necessary reference for theoretical studies and simulations with the aim of understanding the earliest stages of star and planet formation and the time scale of this process.}

\keywords{ISM: clouds - ISM: molecules - radio lines: ISM
               }
\titlerunning{}
\maketitle

%

\section{Introduction}
\label{intro}
Molecular emission is of paramount importance when we try to understand the process of star formation. The wealth of information contained in molecular spectra from star-forming regions encompasses both the physical processes driving the phenomena and the chemical budget available for the forming planets. 
Pre-stellar cores are gravitationally bound objects on the verge of forming a protostar. They represent the initial conditions in stars and planets formation (e.g., \citealt{crapsi07}). Most of the ingredients that will build-up a planet are present already at this stage, from the organic molecules that will form prebiotic material, to the solids that will form the refractory part of the planet \citep{caselli12}. Despite their importance, the dynamical and chemical evolution of pre-stellar cores is still not well understood. Our partial knowledge on the chemical and physical structure of pre-stellar cores, as well as how they are fed and influenced by the surrounding environment, limits the level of knowledge that we can achieve on later stages in the star and planet formation process, from protostellar cores to exoplanets.

This paper is part of a systematic effort aimed at providing a representative sample of dynamically evolved cores that can also be used as targets for higher-resolution studies of the inner core structure, dynamics, and chemistry. The whole sample is presented in Caselli et al. (in prep.). We performed a survey of nearby pre-stellar cores with the Atacama Pathfinder EXperiment (APEX) telescope in rotational lines of N$_2$H$^+$ and N$_2$D$^+$ around 300 GHz and 460 GHz. The pre-stellar cores targeted in this survey represent the densest starless cores within 200 pc from the Sun identified in the $Herschel$ Gould Belt Survey (HGBS, Andre et al. 2010). Through these observations, we want to confirm the high central density derived from $Herschel$ data, and, use the N$_2$D$^+$/N$_2$H$^+$ ratio to estimate their evolutionary stage. With this sample, we are probing the physical conditions immediately before the protostellar birth, necessary to guide theoretical studies and simulations with the overall aim of understanding the earliest stages of star and planet formation and the time scale of this process.
For chemical reasons, N$_2$H$^+$ and N$_2$D$^+$ are indicators of an advanced evolutionary stage of a dense core. Indeed, these molecules form relatively slowly in dense gas from molecular nitrogen, N$_2$, and benefit from the depletion of CO. They, too, deplete in the dense kernel, but the effect is not as severe as for CO and other molecules. Deuterium fractionation is enhanced when CO disappears from the gas phase \citep{roberts03}. As shown by \cite{crapsi05}, a high N$_2$D$^+$/N$_2$H$^+$ abundance ratio ($\geq$ 0.1), is a clear signature of a centrally concentrated core with a very high degree of CO depletion.\\

\begin{figure*}

\begin{center}
\includegraphics[width=18.5cm]{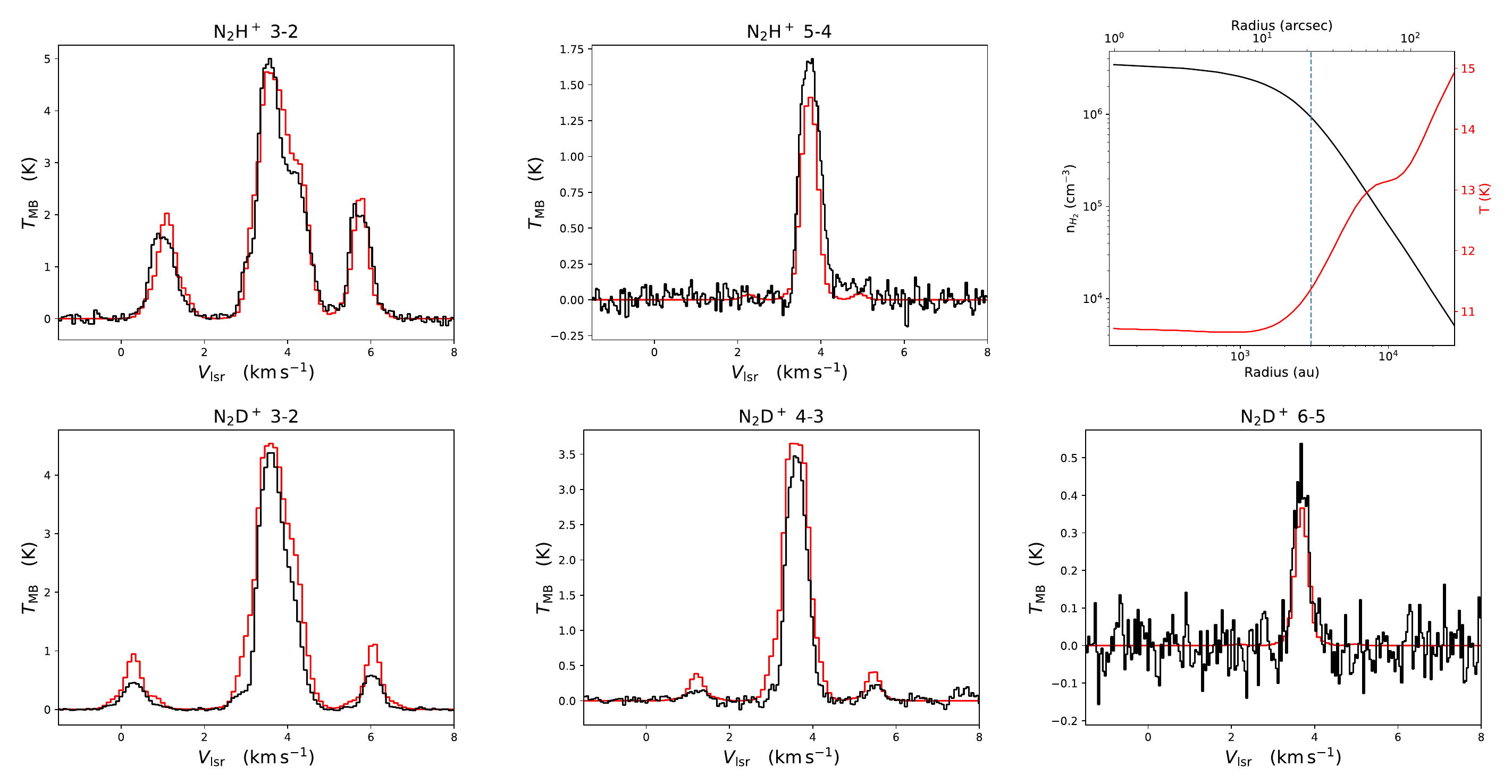}
\end{center}
\caption{APEX observations of several transitions of N$_2$H$^+$ and N$_2$D$^+$ towards the dust peak of the pre-stellar core I16293E (black histograms) and results of the LOC non-LTE radiative transfer model (red curve). The physical model of I16293E computed from $Herschel$ SPIRE and PACS data is shown in the upper right panel. The black curve (left y-axis) shows the total volume density (n$_{H_2}$), the red curve (right y-axis) shows the dust temperature, and the vertical dashed line indicates the 3000 au (21") inner radius of the infalling envelope (see Section~\ref{loc}).}
\label{fig:LOC}
\end{figure*}

This paper focuses on the prestellar core IRAS 16293E (hereafter I16293E), located within the L1689N molecular cloud in the Rho Ophiuchi cloud complex at a distance of $\sim$141pc \citep{dzib18}. I16293E shows some of the largest deuteration levels in the interstellar medium; several multiply deuterated molecules being observed there, including triply deuterated ammonia, ND$_3$ \citep{lis16}. The environment where I16293E is embedded is quite active. The low-mass protostellar system IRAS16923-2422 A and B is located just 90$"$ (12700 au) to the west of the prestellar core \citep{stark04}. Furthermore, there is observational evidence that the large-scale outflows that are driven by the protostellar system have interacted with the pre-stellar core. The area around (and including) I16293E has been extensively studied with both interferometer and single-dish facilities (e.g. \citealt{lis16} and \citealt{kahle23}, and references therein), showing a pattern of velocity that can be explained by the interaction between the prestellar core and the molecular outflow emanating from the nearby solar-type protostars. Through $Herschel$ observations of water and singly deuterated ammonia, \cite{lis16} showed that the outflow wraps around and largely avoids the prestellar core. These authors suggest that the shock associated with the outflow has already propagated through the northeastern part of the core and what can be observed is quiescent, cold, shock-compressed, dense gas, blue-shifted with respect to the systemic velocity of the cloud. The broader line profiles in the western/south-western part of the core suggest instead an ongoing interaction with the outflow towards the south.\\
In this paper, we use a non local-thermodynamic-equilibrium (non-LTE) radiative transfer code to model several rotational lines of N$_2$H$^+$ and N$_2$D$^+$, exhibiting critical densities up to 5$\times$10$^6$ cm$^{-3}$, thus probing the dense central part of the pre-stellar core. Additionally, we investigate the rich chemistry towards I16293E by comparing the line shapes and velocity components in several molecular emission lines of about 20 different species. \\

This paper is structured as follows: in Section 2 we present the observations, the non-LTE modelling of the N$_2$H$^+$ and N$_2$D$^+$ lines is described in Section 3. The line emission observed in several molecules is presented in Section 4. We discuss the results in Section 5 and summarise our conclusions in Section 6.

\begin{table*}[!h]
    \renewcommand{\arraystretch}{1.4}
\centering
\caption{Parameters of the targeted N$_2$H$^+$ and N$_2$D$^+$  lines.}
\label{tab:lines}
\begin{tabular}{cccccccccccc}
       \hline \hline
Transition & Frequency & A$_{ul}$ & g$_{up}$ &n$_{crit}$& E$_{up}$ & $\theta_{MB}$ &  $\eta_{MB}$&$\Delta$ V$_{ch}$ & rms & V$_{lsr}$ & FWHM \\
 		& MHz	& $\times$ 10$^{-3}$/s	&	&cm$^{-3}$	& K		 	&		$"$			&			& km s$^{-1}$ & mK  &km s$^{-1}$ &km s$^{-1}$ \\

\hline
N$_2$H$^+$ $3-2$ &  279511.7491  & $1.26$             &   63 &  1.4$\times$10$^6$   &  27 	& 22.3&	0.76& 0.07     & 43           &  3.610(2)& 0.544(3) \\
N$_2$H$^+$ $5-4$ &  465824.7770 & $6.18$              &    99 &  5.5$\times$10$^6$  &  67 	&13.3	&	0.60	&0.04    & 36     &  3.600(3)& 0.41(1)   \\
N$_2$D$^+$ $3-2$ & 231321.8283  & $0.71$                &  63  & 7.4$\times$10$^5$  &  22 	&26.0	&	0.80	&0.08    & 10      &  3.630(1)& 0.410(1)  \\
N$_2$D$^+$ $4-3$ & 308422.2672  & $1.75$               &   81   & 1.5$\times$10$^6$ &   37	&20.2	&	0.74	& 0.06   & 38     &  3.600(1)& 0.381(7)  \\
N$_2$D$^+$ $6-5$ &  462603.8520 & $6.15$              &    117   & 4.4$\times$10$^6$ &  78 	&13.5	& 0.60		& 0.04 & 35       &  3.620(1)& 0.42(2)    \\

 \hline
\end{tabular}
\tablefoot{ The spectroscopic information refers to the unsplitted rotational transitions and is taken from the CDMS catalogue \citep{muller05}. The last two columns report the kinematic parameters obtained with the \textsc{hfs} fit. The line critical density is computed with Eq. 4 of \cite{Shirley15}, using the collisional rates of the unsplitted lines. \\}
\end{table*}

\section{Observations}
\label{observations}
\subsection{APEX}
\begin{figure*}

\begin{center}
\includegraphics[width=15cm]{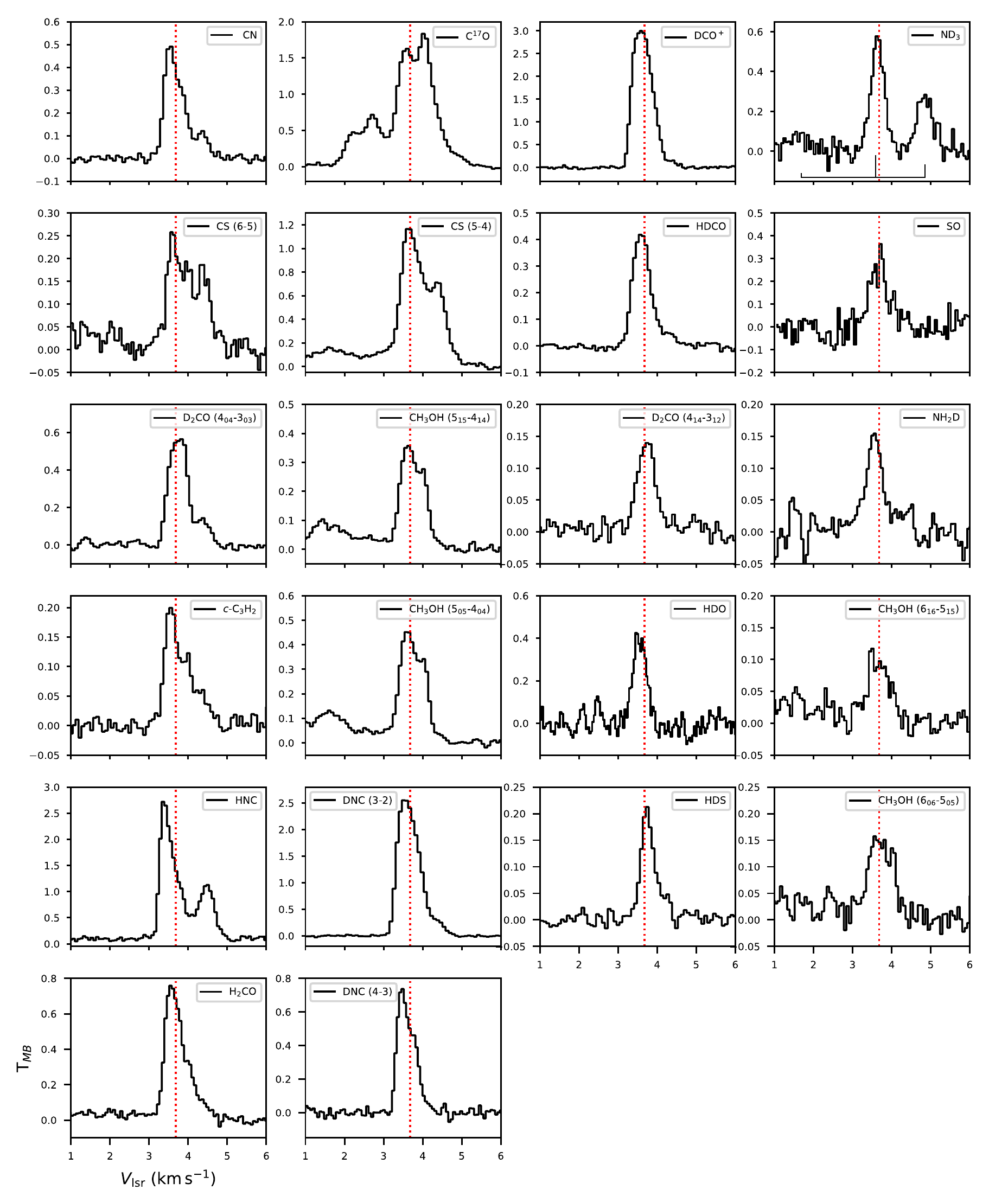}
\end{center}
\caption{Molecular lines observed towards the dust peak of I16293E with APEX. The vertical red dotted line shows the v$_{LSR}$ of the core, 3.7 km s$^{-1}$. Given the complexity of the environment around I16293E, the rest velocity slightly changes depending on the molecule, as already pointed out in \cite{lis16} and \cite{kahle23}. Both lines present in the panel of fully deuterated ammonia (ND$_3$), top right corner, are hyperfine components of the same rotational transition of ND$_3$. The stick spectrum at the bottom of the panel of ND$_3$ shows the hyperfine structure of the
transition. Their spectroscopic parameters are reported in Table~\ref{tab:parameters}.}
\label{fig:all_mol}
\end{figure*}

The spectra observed with APEX towards the dust emission peak of I16293E are shown in Figures~\ref{fig:LOC} and \ref{fig:all_mol}. The molecular emission data presented in this work were collected with the APEX telescope as part of the large core sample described in Caselli et al. (in prep.). 
Two tunings of the SEPIA345 receiver were used to cover the N$_2$H$^+$ 3-2 and N$_2$D$^+$ 4-3 transitions at 279.5$\,$GHz and 308.4$\,$GHz, respectively. These observations were performed in July and October 2022 (Proposal ID: O-0110.F-9310A-2022). The N$_2$D$^+$ 3-2 line was covered with the nFLASH230 receiver, whilst the N$_2$D$^+$ 6-5 and N$_2$H$^+$ 5-4 transitions were covered simultaneously with one tuning of nFLASH460. These observations, collected under project M-0110.F-9501C-2022, were performed in September and November 2022. \\
All observations were performed in ON-OFF position switching (with the OFF position being +212" in R.A and Dec.) toward the core central position ($\rm RA(J2000) = 16^{\rm h}32^{\rm m}29^{\rm s}.07$ and $\rm Dec(J2000)= -24^{\circ}29'09''.0$), consistent with the continuum peak emission at 970 $\mu$m observed with the ALMA compact array \citep{lis16}. We used the FFTS backend with a spectral resolution of 64 kHz. This translates into velocity resolution from $\Delta V_\mathrm{ch}\sim 0.08\,$ km s$^{-1}$ at 230$\,$GHz to $\sim 0.04\,$ km s$^{-1}$ at 460$\,$GHz. The calibration accuracy for the APEX observations is $\sim$10$\%$.\\
 We reduced the data using the CLASS package of the GILDAS software, and subtracted the baseline using a first-order polynomial. The intensity scale was converted into main beam temperature $T_\mathrm{MB}$ assuming the forward efficiency $F_\mathrm{eff} = 0.95$ and computing the main-beam efficiency at each frequency according to $\eta_\mathrm{MB} = 1.00797-0.000857*\nu(\rm GHz)$ \footnote{https://www.apex-telescope.org/telescope/efficiency/index.php}. \\
The spectra in Figures~\ref{fig:LOC} and \ref{fig:all_mol} show the presence of a very rich chemistry as well as a very complex physical structure. In particular, we can see different velocity components in all bright lines, with the exception of the spectra of N$_2$H$^+$ and N$_2$D$^+$ likely because they trace only the pre-stellar core. 
Relevant information on the targeted transitions of N$_2$H$^+$ and N$_2$D$^+$ is given in Table 1.

\subsection{$Herschel$}
\label{herschel}
\begin{figure*}
\unitlength=1mm
\begin{picture}(160,70)(0,0)
\put(142,0){
\begin{picture}(0,0) 
\includegraphics[width=4cm,angle=0]{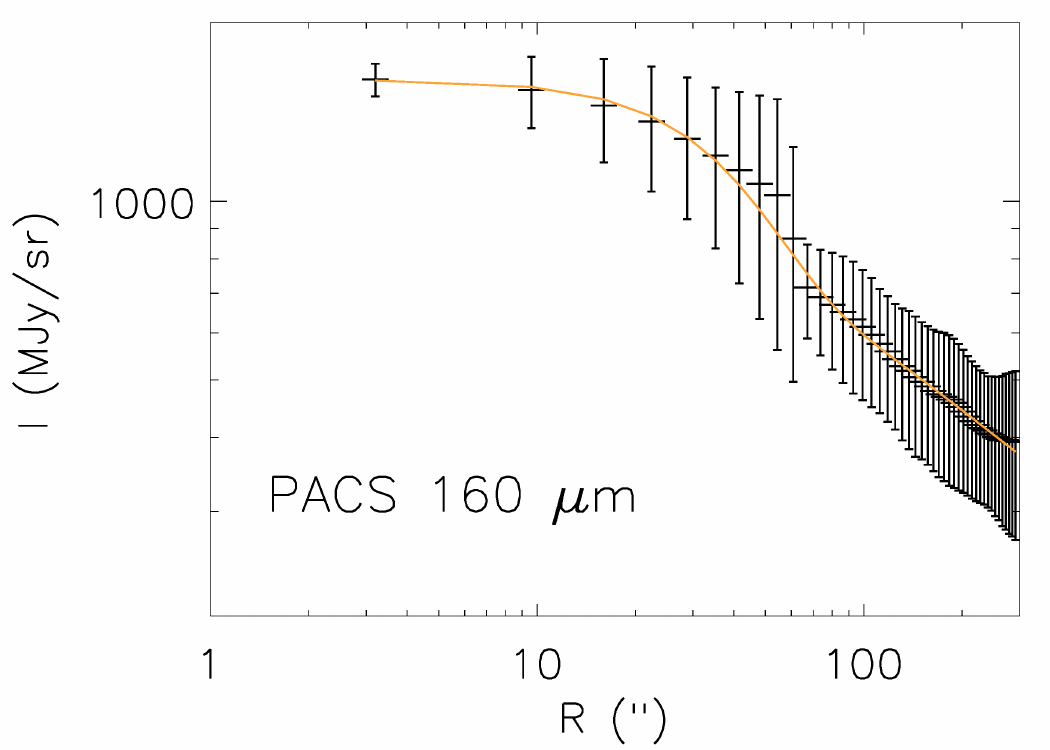}
\end{picture}}
\put(92,0){
\begin{picture}(0,0) 
\includegraphics[width=4cm,angle=0]{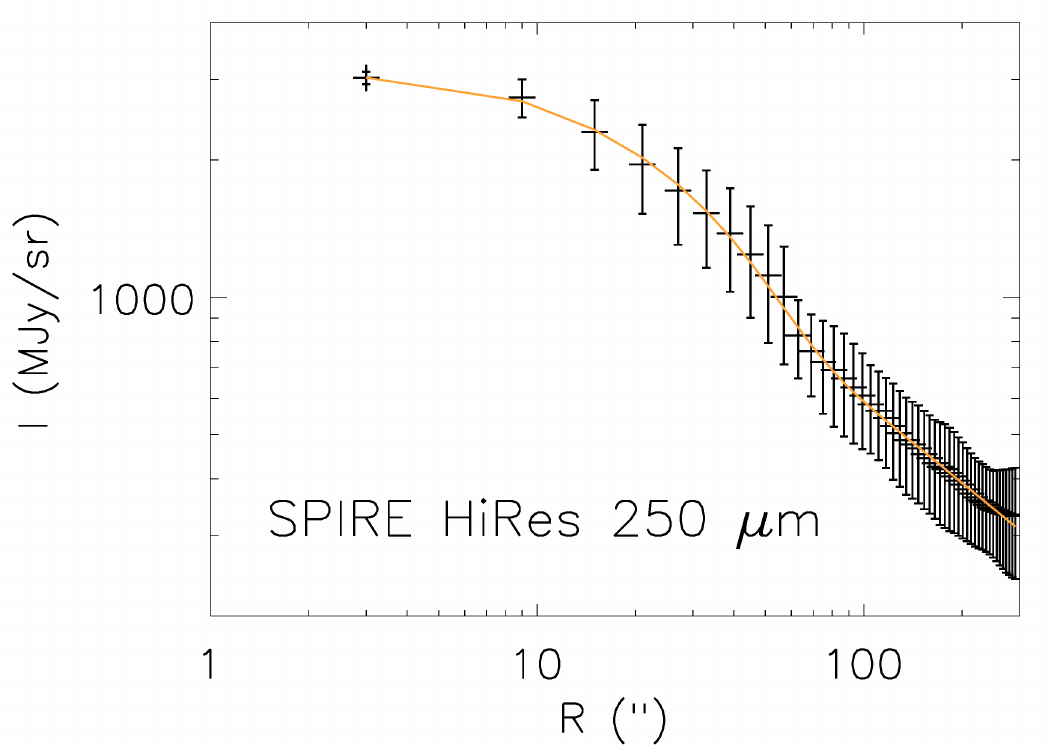}
\end{picture}}
\put(47,0){
\begin{picture}(0,0) 
\includegraphics[width=4cm,angle=0]{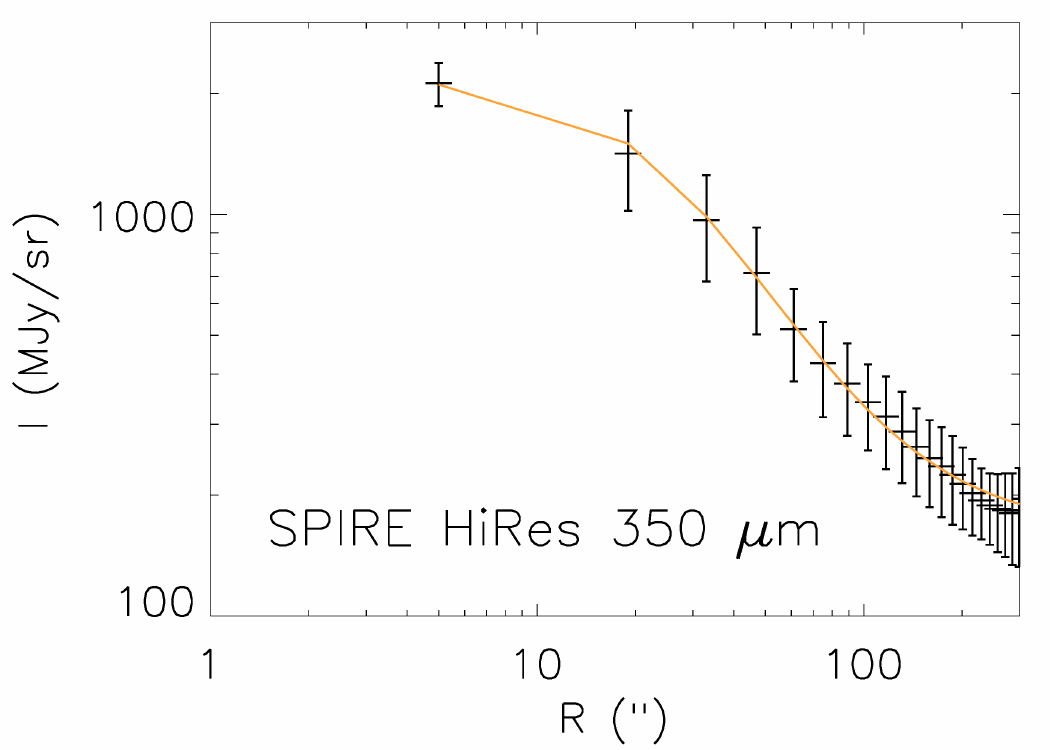}
\end{picture}}
\put(3,0){
\begin{picture}(0,0) 
\includegraphics[width=4cm,angle=0]{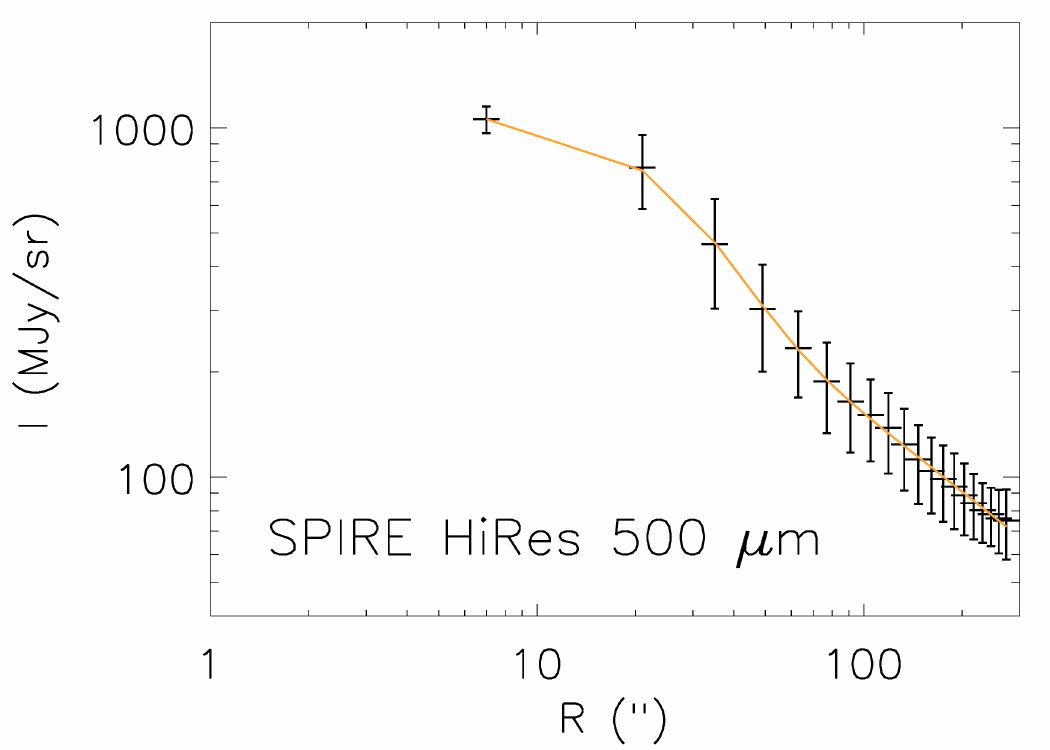}
\end{picture}}
\put(137,30){
\begin{picture}(0,0) 
\includegraphics[width=5cm,angle=0]{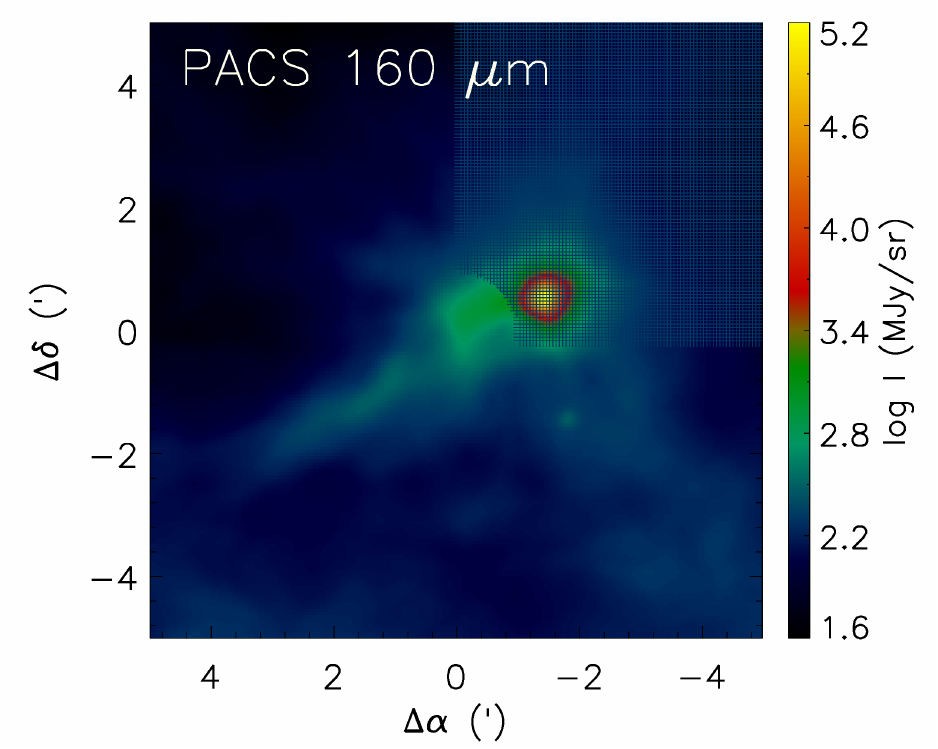}
\end{picture}}
\put(88,30){
\begin{picture}(0,0) 
\includegraphics[width=5cm,angle=0]{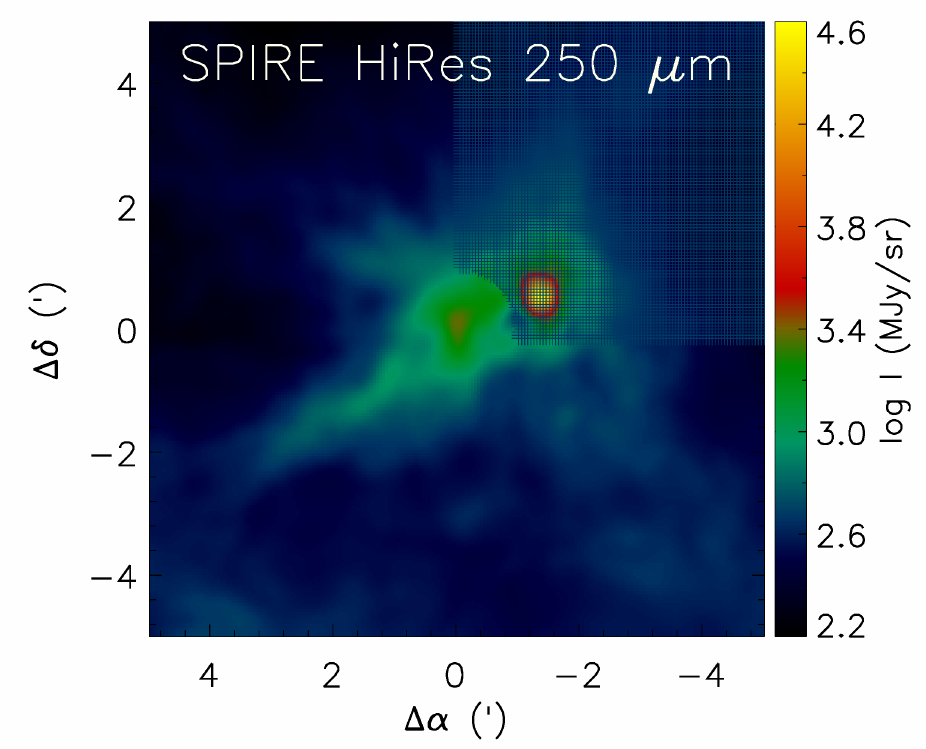}
\end{picture}}
\put(43,30){
\begin{picture}(0,0) 
\includegraphics[width=5cm,angle=0]{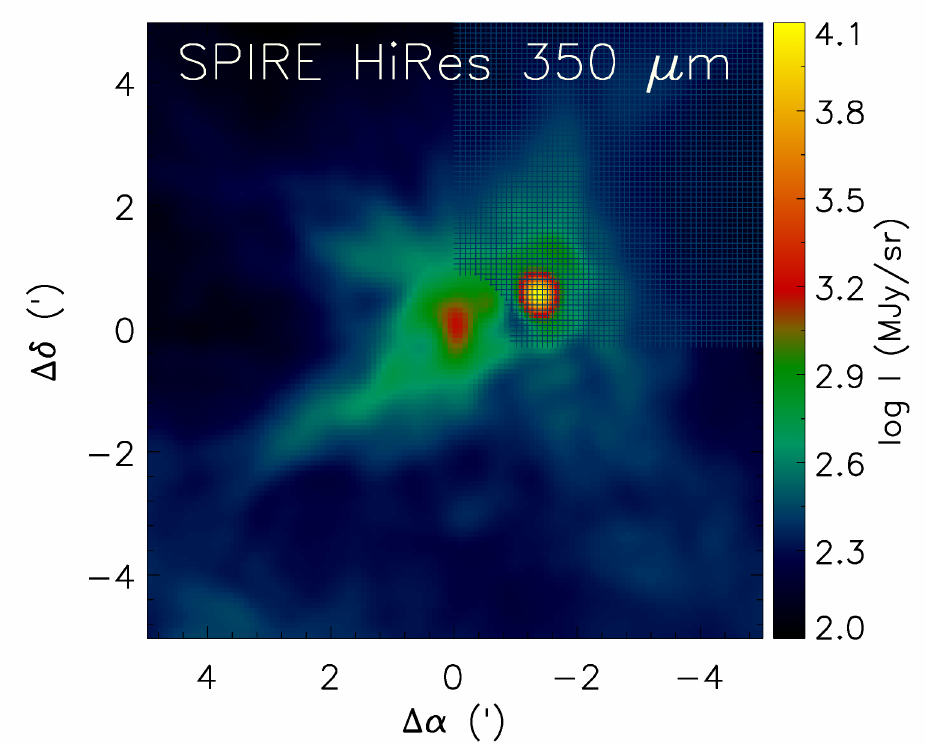}
\end{picture}}
\put(-2,30){
\begin{picture}(0,0) 
\includegraphics[width=5cm,angle=0]{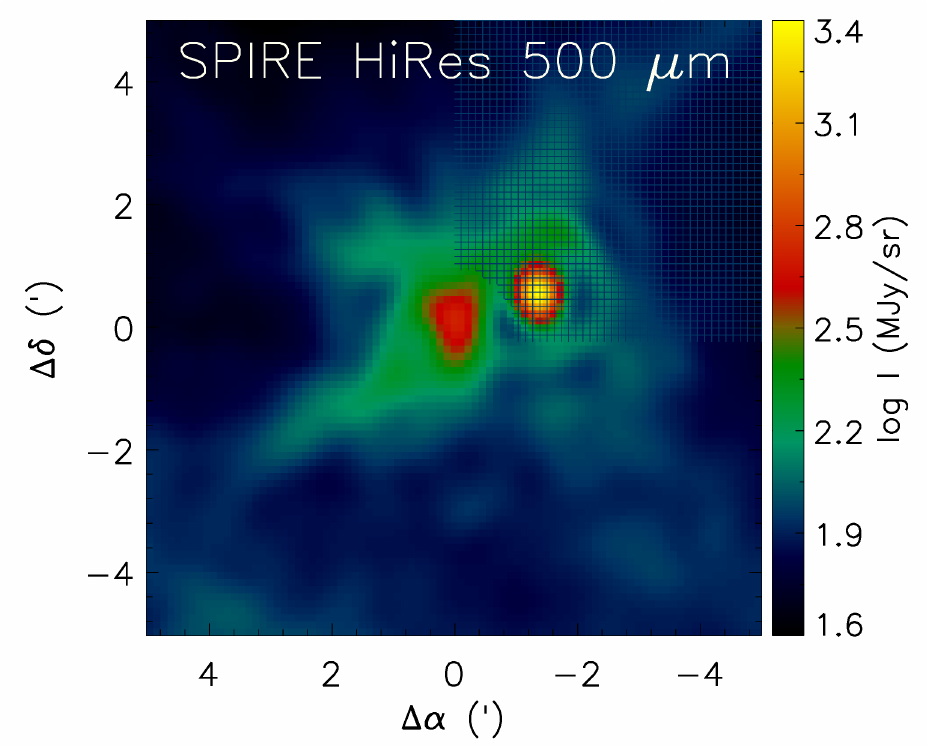}
\end{picture}}

\end{picture}
\caption{Far-infrared maps and averaged intensity profiles of I16293E. {\bf Top:}~ {\sl Herschel} SPIRE HiRes and PACS surface brightness maps at $500\,\mu$m,  $350\,\mu$m, $250\,\mu$m, and $160\,\mu$m. The latter three are smoothed to the SPIRE HiRes resolution $15\arcsec$. The hatched areas are masked out in the calculation of intensity profiles to avoid contribution from IRAS16293-2422. {\bf Bottom:} Circularly averaged surface brightness profiles. The error bars show standard deviations within annular regions. The solid curves show Plummer-type functions fitted to the averages.}
\label{dust_maps}  
\end{figure*}

The top-row panels of Fig.~\ref{dust_maps} show the thermal dust emission maps of the core at four {\sl Herschel} wavelengths. The bottom-row panels of the Figure show the circularly averaged intensity profiles as a function of the impact parameter. The calibration accuracy for SPIRE and PACS have been estimated to be $\sim$15$\%$ \citep{ladjelate20}.
To perform the non-LTE radiative transfer analysis, we computed a spherically symmetric model of the core based on these intensity profiles.  The radial density and dust temperature profiles, $\rho(r)$ and $T_{\rm d}(r)$, of the core model were determined using the method described in \cite{roy14}. According to Eq.\,(3) of \cite{roy14}, the product $\rho(r)\,B_\lambda(T_{\rm d})\,\kappa_\lambda$, where $B_\lambda$ is the Planck function at wavelength $\lambda$ and $\kappa_\lambda$ is the dust opacity at this wavelength, can be obtained from the inverse Abel transformation of the surface brightness gradient $dI_\lambda/dr$ as a function of the impact parameter. The density and temperature profiles were fitted by determining the surface brightness gradient at four wavelengths. For the dust opacity, we assumed $\kappa_{\rm 250 \mu m}  = 0.1 \, \rm cm^{2}\, g^{-1}$ and a dust opacity index of $\beta=2.0$ \citep{hildebrand83}. We used deconvolved {\sl Herschel}/SPIRE HiRes images at $\lambda=250\,\mu$m,  $350\,\mu$m, and $500\,\mu$m (1.2, 0.86, and 0.60 THz, respectively), and the {\sl Herschel}/PACS image of the region at $160\,\mu$m (1.9\,THz).  The SPIRE HiRes images at $500\,\mu$m have a resolution of approximately $15\arcsec$\footnote{http://herschel.esac.esa.int/Docs/SPIRE/spire\_handbook.pdf}. The images at shorter wavelengths were convolved to this resolution. The surface brightness profiles were determined by fitting Plummer-type functions to the concentric circular averages. The purpose of the fitting was to provide a smooth intensity gradient $dI_\lambda/dr$. Figure~\ref{dust_maps} shows the surface brightness maps of a $10\arcmin\times10\arcmin$ region around I16293E. The star-forming core IRAS\,16293-2422 is masked out in the images, and this region was not used in the calculation of the intensity profiles. The error bars in the bottom panel represent the standard deviations of the averages, and the solid curves show Plummer-type fits to the average values. The resulting volume density and dust temperature distributions are shown in the top right panel of Fig.~\ref{fig:LOC}. 
The logarithmic fits of the temperature and density profiles beyond the apparent flat radius at 3000 au (21") until 50000 au (350") are $T\sim r^{0.12}$ and $n\sim r^{-2.45}$. 
The dust temperature ranges from 10.5 to 15 K, slightly warmer compared to typical pre-stellar cores, where the temperatures range from $\sim$6 to $\sim$13 K. The volume density values go from a few times $10^3\,\percc$  in the outer regions to a few $10^6\,\percc$ towards the centre.

\begin{table*}{}
\caption{Spectroscopic parameters of the observed lines}
\label{tab:parameters}
\begin{tabular}{cccccccc}
\hline\hline \\[-2ex]
Molecule & Transition & Rest frequency & $E_\text{up}$ & T$_{A}^*$ & v$_{LSR}$  &   FWHM   & W   \\
&       &(MHz)   & (K)  & K & km s$^{-1}$ & km s$^{-1}$ & K km s$^{-1}$               \\[0.5ex]
\hline \\[-2ex]
C$^{17}$O   & N=2-1 F= 7/2-5/2 and  5/2-3/2  &   224714.19(8)\tablefootmark{a}     &16  & 1.43(1) &3.787(2)& 1.040(4)&1.586(6)   \\
C$^{17}$O   & N=2-1 F= 3/2-3/2 and  5/2-3/2   &    224715.24(8)\tablefootmark{a}     &16  &0.48(1) &3.824(1)& 0.85(1)&0.434(5)   \\
CN    &  N = 2-1 J= 3/2-1/2 F= 1/2-1/2  &  226663.693(2)   & 16 &0.37(8) &3.58(3)& 0.58(9)&0.23(3)   \\
CS    &N= 5-4    &  244935.556(3)   & 35 & 0.81(4) &3.830(9) & 1.13(2)& 0.97(2)  \\
CS    &N= 6-5    &  293912.086(3)   & 49 & 0.16(2) &3.89(2) & 1.12(5)& 0.193(8)  \\
SO    & N J = 2 2 - 1 2   &   309502.44(6)     & 19 & 0.219(3) &3.68(2)& 0.57(5)&0.134(8)  \\
DCO$^+$   & N= 4-3    &  288143.858(3)   & 35 & 3.16(2) &3.589(1) & 0.573(2)& 1.929(6)   \\
HDS    & J$_{K_a K_c}$= 1$_{0 1}$-0$_{0 0}$    &  244555.6(1)   & 12 & 0.20(1) &3.723(7) & 0.44(2)& 0.110(3)   \\
HDO   & J$_{K_a K_c}$= 1$_{0 1}$-0$_{0 0}$  &   464924.5200     &22  & 0.343(3) &3.530(8)& 0.40(2)&0.145(6)   \\
HNC    & J=3-2   &  271981.1420\tablefootmark{b}     & 26 &2.34(2) &3.45(1)& 0.51(3)&1.27(6)   \\
DNC    & J=3-2   &  228910.4810\tablefootmark{b}     & 22 & 2.103(1) &3.600(1)& 0.640(2)&1.433(2)   \\
DNC    & J=4-3   & 305206.22(3)\tablefootmark{b}     & 37 & 0.69(2) &3.534(4)& 0.518(9)&0.380(6)   \\
$o$-NH$_2$D    &  J$_{K_a K_c}$= 2$_{2 1}$-2$_{1 1}$ v$_t$ = 1-0 F = 1-1, 2-2, and 3-3  &  305723.78(3)\tablefootmark{c}     & 72 &0.14(2)&3.63(2)& 0.55(6)&0.083(6)   \\
ND$_3$    & J K F= 1 0 1 - 0 0 1   &   309908.46(10)     &  14& 0.215(3) &3.69(1)& 0.5(1)&0.11(2)   \\
ND$_3$   & J K F= 1 0 2 - 0 0 1   &   309909.69(10)     & 14 & 0.439(3) &3.67(1)& 0.40(3)&0.19(1)   \\
ND$_3$    & J K F= 1 0 0 - 0 0 1   &   309911.53(10)    &  14& 0.05(2) &3.68\tablefootmark{d}& 0.45\tablefootmark{d}&0.02(1)   \\
H$_2$CO   & J$_{K_a K_c}$= 4$_{0 4}$-3$_{0 3}$    & 290623.40(1)   & 35 & 0.72(2) &3.623(4) & 0.62(1)& 0.469(7)   \\
HDCO   & J$_{K_a K_c}$= 4$_{1 3}$-3$_{1 3}$    &  246924.6(1)   & 38 & 0.43(1) &3.563(3) & 0.528(8)& 0.241(3)   \\
D$_2$CO    & J$_{K_a K_c}$= 4$_{0 4}$-3$_{0 3}$  &  231410.23(5)     & 28 &  0.493(1) &3.703(3)& 0.544(8)&0.134(8)   \\
D$_2$CO   & J$_{K_a K_c}$= 4$_{1 3}$-3$_{1 2}$    &  245532.75(5)   & 35 & 0.117(9) &3.723(7) & 0.44(2)& 0.110(3)   \\ 
CH$_3$OH    & J$_{K_a K_c}$= 5$_{1 5}$-4$_{1 4}$    &  241767.234(4)   & 40 & 0.35(4) &3.69(2) & 0.72(4)& 0.27(1)   \\
CH$_3$OH    & J$_{K_a K_c}$= 5$_{0 5}$-4$_{0 4}$    &  241791.352(4)   & 35 & 0.44(4) &3.68(1) & 0.76(3)& 0.36(1)   \\
CH$_3$OH    & J$_{K_a K_c}$= 6$_{1 6}$-5$_{1 5}$    &  290069.747(5)   & 54 &0.10(2) &3.62(3) & 0.70(7)& 0.074(6)   \\
CH$_3$OH    & J$_{K_a K_c}$= 6$_{0 6}$-5$_{0 5}$    &  290110.637(4)   & 49 & 0.12(2) &3.68(2) & 0.78(6)& 0.104(7)   \\
$c$-C$_3$H$_2$    & J$_{K_a K_c}$= 3$_{2 1}$-2$_{1 2}$    &  244222.151(4)   & 18 & 0.133(9) &3.66(1) & 0.77(3)& 0.110(3)   \\
\hline
\end{tabular}
\tablefoot{Numbers in parentheses denote $1\sigma$ uncertainties in unit of the last quoted digit. The frequencies in the Table are reported in the CDMS and JPL catalogues \citep{muller05, pickett98} and are derived from the laboratory work in \citealt{bocquet99, bechtel06, clark76, helminger69, messer84, cazzoli02, klapper03, dixon77, caselli05, helminger71, bruenken06, saykally76, melosso21, zakharenko15, mueller17, xu08, bogey86}.
\tablefoottext{a}{The frequency is averaged between two components, not resolvable in this work, reported in \cite{cazzoli02}. }
\tablefoottext{b}{This frequency does not take into account the hyperfine splitting.}
\tablefoottext{c}{The frequency is averaged between three hyperfine components.}
\tablefoottext{d}{For the fit of this line, the FWHM and v$_{LSR}$ were fixed to the average values of the other two hyperfine components.}
}

\end{table*}

\begin{figure*}

\begin{center}
\includegraphics[width=17cm]{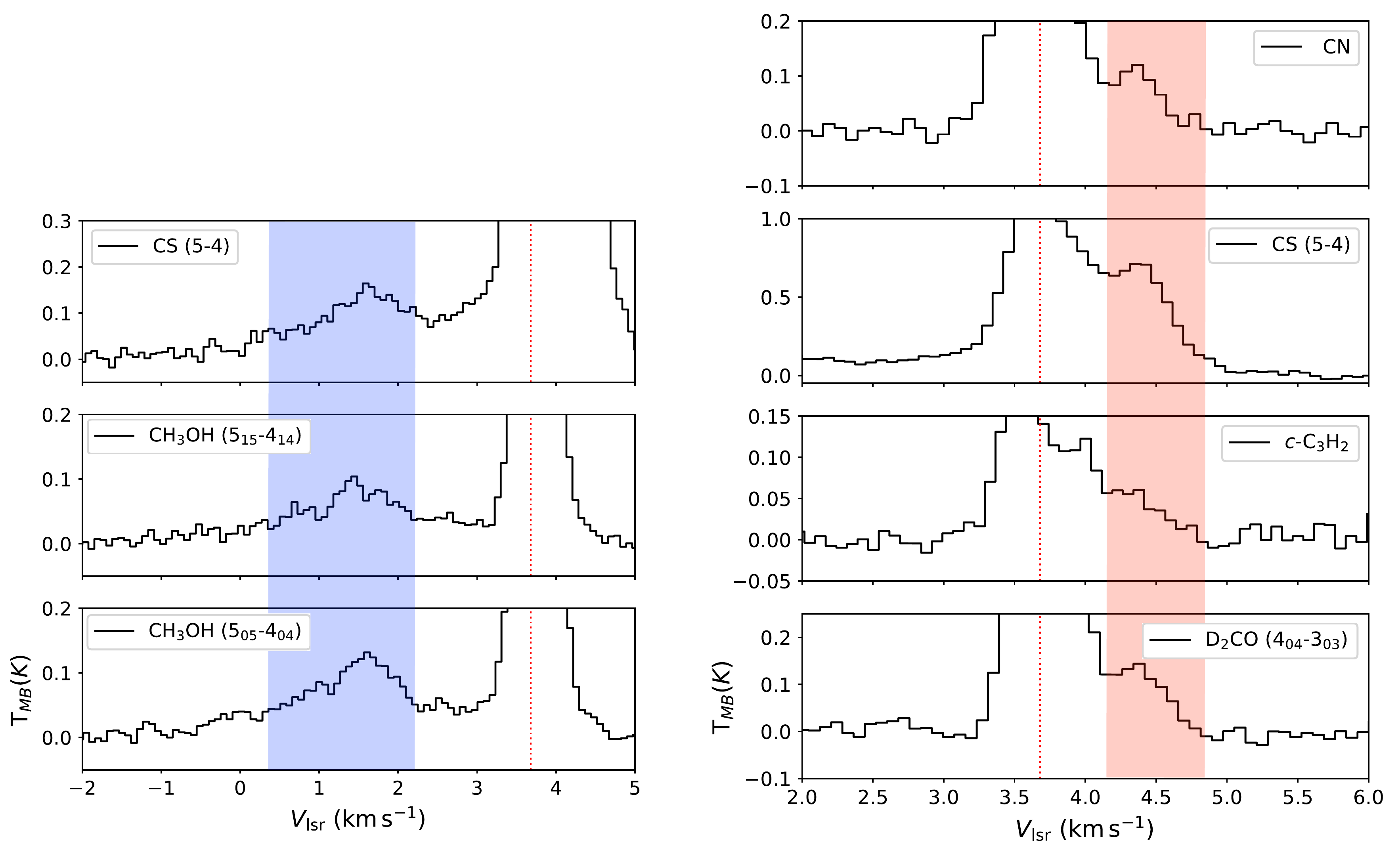}
\end{center}
\caption{Zoom-in of the blue-shifted and red-shifted velocity components present in some of the observed lines. The vertical red dotted line shows the v$_{LSR}$ of the core, 3.7 km s$^{-1}$.}
\label{fig:zoom}
\end{figure*}

\section{Non-LTE modelling of N$_2$H$^+$ and N$_2$D$^+$}
\label{loc}
We used the radiative transfer code LOC \citep{juvela20} to predict the observed N$_2$H$^+$ and N$_2$D$^+$ lines taking into account the overlap of the hyperfine components. As physical models of the source, we used the $\rho(r)$ and $T(r)$ profiles derived from $Herschel$ data as described in Section~\ref{herschel}. We assume T$_{gas}$ = T$_{dust}$ \citep{sokolov17}, where T$_{dust}$ is derived from the modelling of the $Herschel$ data shown in Figure~\ref{dust_maps}. The output spectral cube is convolved with the corresponding observational beam and compared to the observed spectrum. To model our observations we used the hyperfine-resolved collisional coefficients taken from the EMAA catalogue\footnote{Available at \url{https://emaa.osug.fr/species-list}}. The collisional coefficients of N$_2$H$^+$ were computed by \cite{lique15}, and those of N$_2$D$^+$ by \cite{lin20}. 
We tested different abundance profiles for N$_2$H$^+$ and N$_2$D$^+$, and different distributions of the radial velocity and velocity dispersion. 

Simulated spectra from the best-fit model are shown in red in Figure~\ref{fig:LOC}. Here, the turbulent velocity dispersion is set to $\sigma_\mathrm{turb}= 0.195\,$ km s$^{-1}$, and the radial velocity profile is a step function where the velocity is set to zero in the inner 3000 au (21") and -0.5 km s$^{-1}$ at larger radii, i.e. a static inner core surrounded by infalling envelope. Also the abundance profiles are step functions with values of 4.5$\times$10$^{-10}$ for N$_2$H$^+$ and 2$\times$10$^{-10}$ for N$_2$D$^+$ in the inner 3000 au (21") and a drop of 70\% of the central values at radii larger than 3000 au.\\

\textbf{Velocity profile} To reproduce the asymmetric line profile of the N$_2$H$^+$ 3-2 line, it is necessary to use a velocity radial profile with a velocity of -0.5 km s$^{-1}$ at radii larger than 3000 au (21") and equal to zero in the central part of the core. These values have been determined by testing a range of typical values for similar objects. A negative velocity means that the gas is moving towards the centre of the core. Such infalling motions are responsible for red-shifted absorption. 
While it is possible to reproduce well the high-J transitions with a static core (i.e. v=0 in the whole core), the asymmetry in the  N$_2$H$^+$ 3-2 central hyperfine transition can only be reproduced with an infall velocity profile. The results of our test with a static core are shown as a green curve in Figure~\ref{fig:vel_tests}. We have also tried to "invert" the step function of our best-model velocity profile, assuming our core to have an infalling central region and a static envelope. In this case the velocity was set to -0.5 km s$^{-1}$ in the central part of the core and equal to zero at radii larger than 3000 au (21"). The results of this test are shown as a blue curve in Figure~\ref{fig:vel_tests}. In this case neither the low-$J$ nor the high-$J$ transitions are well reproduced. In particular, the high-$J$ transitions of both N$_2$H$^+$ and N$_2$D$^+$ appear to have separated velocity components because of the infall velocity, which is not the case for the observed spectra.\\
 A radius of 3000 au (21") for the static core proves to be a good value to reproduce the APEX observations. When decreasing the radius to 1000 au (7"), the intensity of the high-J transitions decreases significantly as the lines become broader because of the effect of the infall. If we instead increase the radius of the static core to 4000 au (28"), the asymmetry in the N$_2$H$^+$ 3-2 line decreases and disappears with further increasing of this value. We note that a step function is obviously a very crude description of the velocity profile of I16293E. Nevertheless, our results show how powerful multi-line non-LTE modelling of N$_2$H$^+$ and N$_2$D$^+$ is to constrain the physical and dynamical structure of a pre-stellar core. \\

\textbf{Abundance profile} Our N$_2$H$^+$ and N$_2$D$^+$ observations can be well reproduced using a step-function abundance profile with a drop of 70\% at radii larger than 3000 au (21"). The central abundances that best reproduce our observations are 4.5$\times$10$^{-10}$ and 2$\times$10$^{-10}$ for N$_2$H$^+$ and N$_2$D$^+$, respectively, one of the largest D/H ratio in N$_2$H$^+$ observed to date (0.44). The central regions of pre-stellar cores are characterised by very high density (on the order of few 10$^6$ cm$^{-3}$) and very low temperatures (down to $\sim$6 K). One of the consequences of such low temperatures and high densities is the freeze-out of molecules on top of the icy mantles of dust grains in the central regions of the cores, and hence a depletion of molecules from the gas phase. Recent ALMA observations of deuterated ammonia in the low-mass pre-stellar core L1544 in the Taurus molecular cloud showed an almost complete freeze-our in the inner 1800 au of the core, where $\geq$99\% of the gas is frozen on the dust grains \citep{caselli22, pineda22}. 
To examine whether the same effect is observable in I16293E, we ran models with an abundance drop in the central part of the core. It turned out that such a drop would dramatically weaken the high-J lines of N$_2$H$^+$ and N$_2$D$^+$, and that models with a constant abundance in the core nucleus reproduce best the observations. Scaling up the fractional abundances but keeping the centrally decreasing profiles would not help, because the high-J lines would still be too weak and the low-J lines would become optically thick. The results of our tests using different depletion radii are reported in the Appendix. A complete depletion in the inner 2000 au (14") of the core significantly affects the high-$J$ lines of both molecules, see the blue curve in Figure~\ref{fig:depletion_tests}. The high-$J$ lines are much better reproduced if we consider a complete depletion in the inner 1000 au (7") of the core, as can be seen in Figure~\ref{fig:depletion_tests}, green curve. 
The spatial resolution of the present observations is approximately 2000 au (14"), meaning that observations with higher angular resolution are necessary to make a conclusive statement about the freeze out of N$_2$H$^+$ and N$_2$D$^+$ in the centre of I16293E. We note that the weaker hyperfine components of the 3-2 and 4-3 lines of N$_2$D$^+$ are overestimated by our best model shown in Figure 1. While scaling up the volume density profile and simultaneously decreasing the molecular abundance would help in this regard, we prefer to keep the physical structure of the core derived from the $Herschel$ data fixed.
\cite{daniel16} derived an abundance profile for N$_2$H$^+$ by modelling the maps of the 1-0 and 3-2 transitions. The abundance they derive at radii larger than 15" ($\sim$2000 au) is 3-4$\times$10$^{-10}$ and it agrees quite well with the abundance derived in this work (3-4.5$\times$10$^{-10}$). \cite{lis02} showed that the emission of some simple hydrogenated and deuterated molecules in L1689N (the cloud where I16293E is embedded) comes from distinct physical regions. \cite{bacmann16} showed that the ND emission and the NH absorption in I16293E originate from different layers in the cloud. However, the authors point out that this might only be a consequence of the difference in excitation for the two species and that the molecules are both present in the inner dense region of the core, while only NH is present in the outer region.  Our work here confirms this, as the emission of N$_2$H$^+$ and N$_2$D$^+$ is very well modeled assuming that the two molecules are co-spatial. As dissociative recombination of N$_2$H$^+$ and N$_2$D$^+$ with electrons is the main pathway to form NH and ND \citep{roueff15}, we can safely assume that also NH and ND will be present in the same gas in I16293E in the inner dense region of the core. This is also confirmed by the abundance profiles from the chemical model described in \cite{sipila19}, computed at 1$\times$10$^5$ yr, as shown in Figure~\ref{fig:profiles}.

As discussed for the velocity profile, using a step function for the abundance profile is also a crude approximation of the true distribution of N$_2$H$^+$ and N$_2$D$^+$ across I16293E. Thus, we have also used a different approach and coupled the physical structure of I16293E derived from the $Herschel$ observations (cf. Section~\ref{observations}) with the chemical model presented in \cite{sipila19}. The results of the LOC modelling using the abundance profiles from the chemical models are not satisfactory and in particular cannot reproduce the high-$J$ lines of neither N$_2$H$^+$ nor N$_2$D$^+$, see Figure ~\ref{fig:olli}. The models in fact predict a severe depletion towards the inner part of the core starting already at radii larger than 5000 au (45"). Because the high-$J$ transitions have high critical densities, the depletion in the central part of the core has a severe effect on the amount of gas that is traced by the N$_2$H$^+$ 5-4 and N$_2$D$^+$ 6-5 lines. 
The results of the modelling using the abundance profiles computed at 1$\times$10$^5$ yr are shown in Figure~\ref{fig:olli} in blue. Abundance profiles extracted at later times are more affected by the depletion in the central part of the core and consequently reproduce even worse the observed high-$J$ lines. 
To decrease the effect of the depletion in the inner part of the core, we tested the effects of a different prescription for the cosmic ray ionisation rate ($\zeta$), described in \cite{padovani18}.
A higher $\zeta$ could be justified by the proximity to an active site of star formation, as energetic particles can be accelerated along the outflows driven by young stellar objects \citep{padovani16}. Additionally, a higher $\zeta$ is expected to keep more N$_2$ in the gas phase in the central part of the core.
We used two models with a radial dependence of $\zeta$, one with the high model ($\mathscr{H}$) and one with the low model ($\mathscr{L}$) from \cite{padovani18}, as done in \cite{redaelli21} to model the emission of N$_2$H$^+$ in the pre-stellar core L1544. The results of the modelling using these abundances are shown in Figure~\ref{fig:olli}, in green ($\mathscr{H}$) and red ($\mathscr{L}$).
These models did not translate in a big improvement on reproducing the high-$J$ lines because the depletion is still significant at radii larger than 3000 au. 
Additionally, these models predict N$_2$H$^+$ and N$_2$D$^+$ to be abundant in the outer parts of the core, resulting in overly bright (and broad) low-J lines of these molecules. Further studies on the effect of an enhanced cosmic ray ionisation rate on the abundance of N$_2$H$^+$ and N$_2$D$^+$ are necessary to assess the effect of the environment on the depletion radius of these two molecules. 
More details on our tests with the chemical models are presented in the Appendix.

\section{Molecular emission towards the dust peak of I16293E}
While the main objective of the present observations was to use N$_2$H$^+$ and N$_2$D$^+$ lines for confirming the high central density of the core, the large bandwidths of the APEX receivers allowed us to simultaneously detect several additional molecular lines. These are listed in Table~\ref{tab:parameters} and shown in Figure~\ref{fig:all_mol}. The results of the Gaussian fits (performed with the \textsc{class} software) on all observed lines are also presented in Table~\ref{tab:parameters}. We observed both simple species like diatomics and triatomics (C$^{17}$O, CS, CN, SO, HDO, HDS, DCO$^+$, HNC, and DNC), as well as larger molecules (H$_2$CO, HDCO, D$_2$CO, CH$_3$OH, $c$-C$_3$H$_2$, NH$_2$D, and ND$_3$). Among the species that we observe, deuterated molecules are strikingly dominant. Half of the molecules reported in Table~\ref{tab:parameters} are in fact deuterated, and we also observe relatively bright lines of a triply deuterated species, ND$_3$.\\
Some of the lines shown in Figure~\ref{fig:all_mol} have a single Gaussian profiles and do not show hints of multiple velocity components. This is the case for HDO, HDS, NH$_2$D and ND$_3$. Other lines instead show multiple velocity components. The brightest methanol and CS lines have a quite broad ($\sim$1.5 km s$^{-1}$) velocity component blue-shifted from the v$_{LSR}$, between 0.5 and 2 km s$^{-1}$. The lines of CS, CN, $c$-C$_3$H$_2$ and D$_2$CO instead have an additional component redshifted with respect to the core v$_{LSR}$, at $\sim$4.5 km s$^{-1}$. Both blue-shifted and red-shifted velocity components are shown in Figure~\ref{fig:zoom}. In order to understand where these velocity components might originate from, it is necessary to inspect the crowded area surrounding the pre-stellar core.
Maps of over 30 molecules in the area around I16293E have been recently reported by \cite{kahle23}, with an exhaustive study of the large scale molecular emission in the pre-stellar core, the envelope surrounding the IRAS 16293-2422 A and B protostars, and the shocked dense gas in the area. 
The CO $J$=6-5 emission map shown in Figure 2 of \cite{kahle23} is dominated by the presence of two outflows that originate from the IRAS 16293-2422 A and B protostars and that interact with the pre-stellar core. In particular, the blue side of the east-west outflow broadens around the I16293E pre-stellar core, appearing to wrap around the dense core structure. The red side of the northeast-southwest outflow instead contributes to two emission peaks of shocked dense gas in the north of I16293E, E1 and E2. 
Given their velocity and vicinity to the core, we suggest that the red- and blue-shifted components observed in some of the emission lines towards the dust peak of I16293E trace gas from outflows that generated the E1 and HE2 emission peaks. Such gas is not part of the pre-stellar core, but it is in the line of sight. 

\section{Discussion: a dense pre-stellar core in an active environment}
The non-LTE radiative transfer modelling of the N$_2$H$^+$ and N$_2$D$^+$ lines showed how sensitive these lines are to small changes in the velocity and abundance profiles, and have proved to be excellent diagnostics for the gas they trace. To reproduce the asymmetries observed in the low-$J$ transitions, the gas needed to be infalling. The best velocity profile to reproduce all observed lines of N$_2$H$^+$ and N$_2$D$^+$ is a static core in the inner 3000 au (21") surrounded by an infalling gas layer.
The velocity difference between the inner static core and the envelope in our model (0.5 km s$^{-1}$) is consistent with the ND/NH observations in \cite{bacmann16}, where NH absorption is red-shifted with respect to ND emission by $\sim$0.4 km s$^{-1}$.\\
The N$_2$H$^+$ and N$_2$D$^+$ lines proved to be very sensitive to depletion, in particular the high-$J$ transitions. As discussed in Section~3, an abundance profile with no depletion in the centre reproduces our observations best. We also get good results with a depletion radius of 1000 au (7"), which is however a spatial scale that we cannot resolve with our $Herschel$ observations, and hence in the physical structure that we use for the modelling. Our results using a depletion radius of 1000 au (7") are in agreement with the very low abundance of N$_2$D$^+$ reported in \cite{lis16} with ALMA observations within a 1300 au (9") beam.
Testing the abundance profiles predicted by chemical models using the physical structure of I16293E gave us interesting insights on how to further develop the chemical models. The chemical model that we used \citep{sipila19} predicts too much depletion and consequently cannot reproduce the high-$J$ lines of N$_2$H$^+$ and N$_2$D$^+$. We also used two models with a radial dependence of the cosmic ray ionisation rate, following the work of \cite{padovani18}, and the one with the high cosmic ray ionisation rate ($\mathscr{H}$) does a good job reproducing the high-$J$ lines, but not the low-$J$ ones. A higher cosmic ray ionisation rate keeps more molecular nitrogen, N$_2$ in the gas phase in the central part of the core, but it also predicts high abundances of  N$_2$H$^+$ and N$_2$D$^+$ in the outer parts of the core, see Figure~\ref{fig:profiles}. The presence of a considerable amount of  N$_2$H$^+$ and N$_2$D$^+$ in the outer gas layers, where the gas temperature is 13-15 K, translates into an additional broader line component in the N$_2$H$^+$ and N$_2$D$^+$ 3-2 lines (quite noticeable in the green and red curves in the left panels in Figure~\ref{fig:olli}), which is not consistent with the observed spectra. \\
To better reproduce the observed lines in a pre-stellar core like I16293E, we plan to include the effects of the shock in the chemical models. A realistic model of I16293E should also combine chemistry with the dynamic evolution of the core.
In particular, I16293E has likely evolved fast into its present high-density state (due to e.g. shock compression), which is not captured by the present model and hence the depletion is overestimated.\\ 
The N$_2$D$^+$/N$_2$H$^+$ ratio that best reproduces our observations is 0.44, one of the largest observed to date in pre-stellar cores. Using the LTE-approach, \cite{kahle23} derived a ratio of $\sim$10$\%$ (Table F.2 in their paper). The large discrepancy between the two values strongly suggests that non-LTE effects have quite an impact on these spectra. 
I16293E is well-known to have some of the largest deuteration fractions among pre-stellar cores \citep{lis16}. How much of the observed deuteration is a consequence of the environment where the core is embedded still needs to be explored in a quantitative manner. 
It is important to note that half of the molecules that we observed in our spectra are deuterated isotopologues, confirming the high-level of overall deuteration towards this source. A quantitative study of the deuteration of different classes of molecules will be relevant but it is beyond the scope of this paper. Our results on N$_2$H$^+$ show that the complex morphology of the source warrants synergetic approaches that include observations, radiative transfer modelling and chemical modelling, when possible. 
The presence of line components red-shifted and/or blue-shifted from the rest velocity of the core is a consequence of the complex morphology of the source as discussed in Section 4.
 We note that the N$_2$D$^+$ and N$_2$H$^+$ lines are the only bright lines in our sample that show no additional velocity components, thus proving to be excellent tracers of the dense core also in the case of a crowded environment like the one where I16293E is embedded. \\
In a recent paper, \cite{pagani24} present para-$\dtwohplus$ and ortho-$\htwodplus$ maps of the IRAS16293E core. The emission of both molecular ions peak approximately $25\arcsec$ north-east from the dust emission peak (our centre position). Because $\htwodplus$ and $\dtwohplus$ are predicted to be most abundant in cold, dense, and heavily depleted gas, Pagani et al. suggest that the true centre of the core lies at the $\dtwohplus$ peak, and that the dust emission peak, which roughly coincides with the $\diaz(4-3)$ and $\rm ND_3(1_0-0_0)$ emission peaks as observed with ALMA \citep{lis16} indicates a "hotspot" resulting from compression by outflows from IRAS 16293-2422. In Fig.~\ref{tc_and_col_maps} we show the dust colour temperature ($T_{\rm C}$)  and $\htwo$ column density ($N(\htwo)$) maps of the region, derived from {\sl Herschel} PACS  and SPIRE/HiRes images at the wavelengths $\lambda=70$, 100, 160, 250, 350, and $500\,\micron$. The maps are obtained by fitting a grey body function pixel-by-pixel to aligned surface brightness maps smoothed to a common resolution, with the $T_{\rm C}$ and $N(\htwo)$ as free parameters. The $T_{\rm C}$ represents the line-of-sight average of the dust temperature. Both the dust emission peak and the $\dtwohplus$ peak, indicated with the diamond in the maps, are found in a valley of relatively low colour temperatures ($T_{\rm C}< 13$\,K), where the $\htwo$ column densities exceed $10^{23}\,\persqcm$. The $N(\htwo)$ at the $\dtwohplus$ peak is $1.5\times10^{23}\,\persqcm$, which agrees with the core model of \cite{pagani24}. The $\htwo$ column density reaches its maximum at the centre position adopted here. This is no surprise, because the position was chosen from {\sl Herschel} HiRes column density maps provided by the HGBS archive. The $N(\htwo)$ and $T_{\rm C}$ distributions derived from {\sl Herschel} reproduce very well the $850\,\micron$ surface brightness map of the core observed with SCUBA (\citealt{pattle15}), which also peaks at our (0,0). This position probably represents a true maximum in the total column density. Usually the highest densities are found in the direction of $N(\htwo)$ maxima, and probably also here it contains a local density maximum. The present line observations, in particular the strong detection of the $\diaz(5-4)$ and $\ddiaz(6-5)$ transitions confirm that the density in this position exceeds $10^6\,\percc$, which agrees with the estimate from the Abel transformation method described in Sect. 2.2.  We note that according to the 1D physical model used by \cite{pagani24}, the volume density at the offset $25\arcsec$ from the $\dtwohplus$ peak is $\sim 4\times 10^5\,\percc$. This density would be too low to produce the observed intensities of $\diaz(5-4)$ and $\ddiaz(6-5)$ in this position. However, according to our tests, another core resembling the inner portions ($r<10^{17}$\,cm) of the model of \cite{pagani24} (see their Fig.~4) with a central temperature of 8\,K can be placed in the direction of the $\dtwohplus$ peak without substantial changes in the observable $850\,\micron$ map.

\section{Summary and Conclusions}
Several rotational transitions of N$_2$H$^+$ and N$_2$D$^+$ have been observed towards the dust peak of the pre-stellar core I16293E using APEX, namely the N$_2$H$^+$ 3-2 and 5-4 and the N$_2$D$^+$ 3-2, 4-3 and 6-5 transitions.  
I16293E is a dense pre-stellar core located close to the well-studied low-mass binary protostar IRAS 16293–2422. The physical structure of I16293E, constrained from $Herschel$ continuum maps, shows very high central density ($\sim$3$\times$10$^6$ cm$^{-3}$) and temperatures ranging from 11 to 15 K.
The best model to reproduce our spectra of N$_2$H$^+$ and N$_2$D$^+$ hints at the presence of a static central core ($\sim$3000 au, 21") surrounded by an infalling gas layer. Furthermore, the high-$J$ N$_2$H$^+$ and N$_2$D$^+$ lines that we observe are very sensitive to depletion because of their high critical densities (up to$\sim$5$\times$10$^6$ cm$^{-3}$) and our best model does not include any depletion towards the centre of the core. Because I16293E is embedded in a very active environment, it might undergo a local enhancement of cosmic rays that could partially counteract high densities and low temperatures that cause molecular depletion towards the centre of pre-stellar cores \citep{ceccarelli14, padovani16}. Additionally, because of the possible passage of a shock \citep{lis02, lis16}, the high-density nucleus in I16293E may have formed quickly, leaving not enough time for substantial freeze-out.
Our tests cannot completely exclude depletion, especially at radii $<$1000 au (7"). High-angular resolution observations to derive the physical structure of the inner part of the pre-stellar core will be necessary in this regard.
We could not reproduce the N$_2$H$^+$ and N$_2$D$^+$lines using abundance profiles derived from chemical models. Nevertheless, the insights gained from these tests will be used to further constrain our models, especially concerning the radial dependence of the cosmic ray ionisation rate. \\
From the abundance profiles of our best models for N$_2$H$^+$ and N$_2$D$^+$, we derive a deuterium fraction of 0.44, among the largest values measured in similar objects. \\

The wide frequency coverage of the APEX data allowed us to detect the emission of 16 additional molecules, half of them being deuterated isotopologues. 
Given the complex morphology of the area, and the limited angular resolution of our observations (13$"$-26$"$), we used the line profiles to disentangle different gas components in the line of sight. Some lines show multiple velocity components and we suggest that some shocked gas from the outflows originating from the IRAS 16293-2422 A and B protostars contribute to the emission of molecules like CS and CH$_3$OH. This is very clearly seen in the brightest lines shown in Figure~\ref{fig:all_mol}. 
There are however some lines that show a single Gaussian profiles and no hints of multiple velocity components. This is the case for HDO, HDS, NH$_2$D and ND$_3$, as well as N$_2$H$^+$ and N$_2$D$^+$. \\
High spectral resolution observations coupled with radiative transfer modelling and supported by state-of the art chemical models are mandatory to dissect the contribution of different gas components towards I16293E, and finally determine the chemical ingredients that are available for the future planetary systems that will form in the area. Future studies coupling observations and modelling should in particular explore how the enhanced deuteration affects different classes of molecules, with a particular focus on complex organic molecules.

\begin{acknowledgements}
The authors wish to thank the anonymous referee for the careful review of the manuscript.
     S.S, E.R, P.C. O.S, J.E.P. gratefully acknowledge the support of the Max Planck Society. 
      The data was collected under the Atacama Pathfinder EXperiment (APEX) 
Project, led by the Max Planck Institute for Radio Astronomy at the ESO La 
Silla Paranal Observatory. 
     This research has made use of data from the Herschel Gould Belt survey (HGBS) project (\url{http://gouldbelt-herschel.cea.fr}). The HGBS is a Herschel Key Programme jointly carried out by SPIRE Specialist Astronomy Group 3 (SAG 3), scientists of several institutes in the PACS Consortium (CEA Saclay, INAF-IFSI Rome and INAF-Arcetri, KU Leuven, MPIA Heidelberg), and scientists of the Herschel Science Center (HSC). This research has made use of spectroscopic and collisional data from the EMAA database (https://emaa.osug.fr and https://dx.doi.org/10.17178/EMAA). EMAA is supported by the Observatoire des Sciences de l’Univers de Grenoble (OSUG).

\end{acknowledgements}

%
%

{}

\begin{appendix}
\label{appendix}
\onecolumn

\section{Velocity profile and abundance profile tests}
Different velocity and abundance profiles have been tested to reproduce the N$_2$H$^+$ and N$_2$D$^+$ lines using the physical structure derived from $Herschel$ observations (see Section~\ref{herschel}).

\begin{figure*}

\begin{center}
\includegraphics[width=18.5cm]{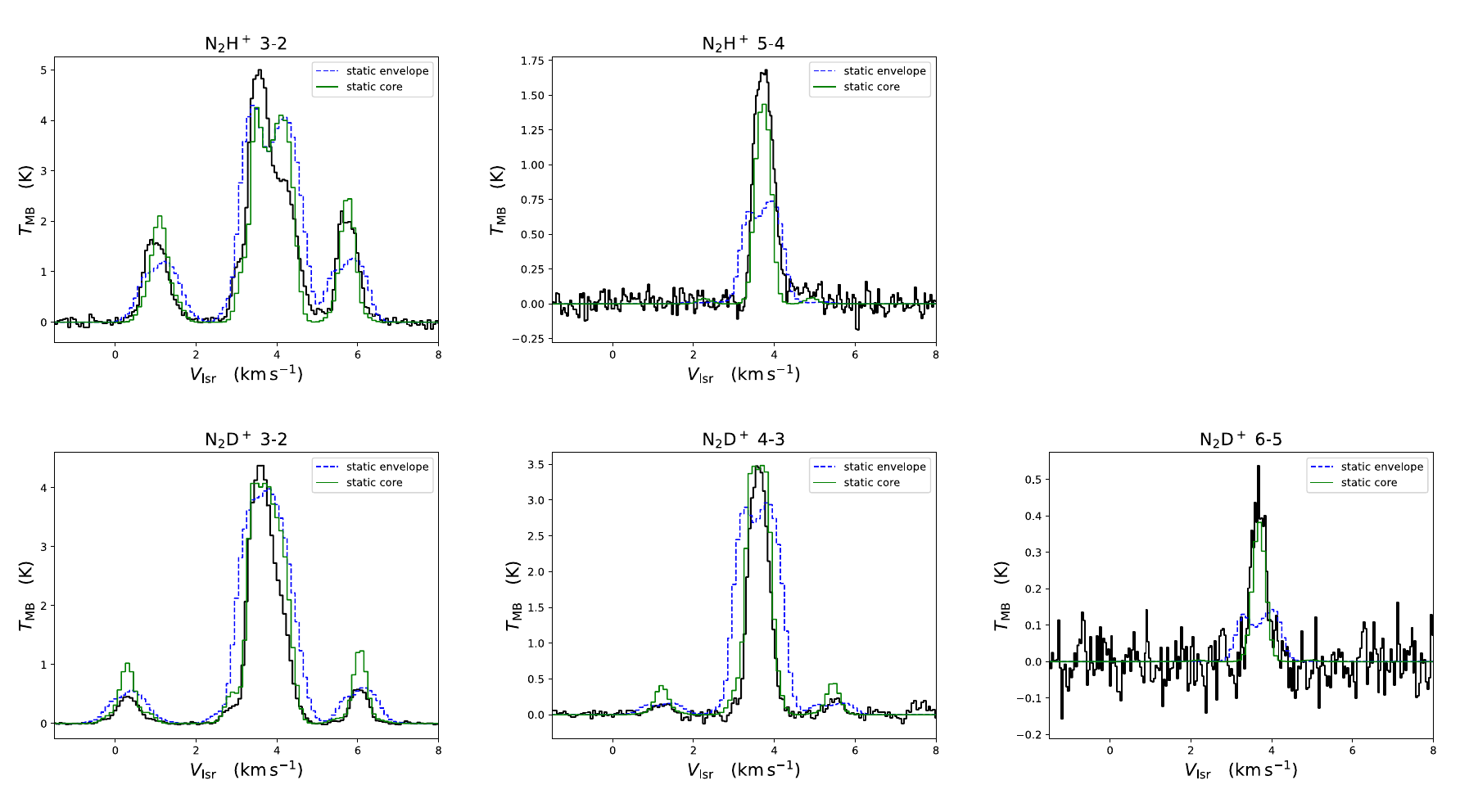}
\end{center}
\caption{Results of tests for the velocity radial profile. The green curve shows the results of the non-LTE modelling done assuming a static core (i.e. v=0 in the whole core). The blue dashed curve shows the results of the non-LTE modelling done assuming a static envelope (the velocity was set to -0.5 km s$^{-1}$ in the central part of the core and equal to zero at radii larger than 3000 au.}
\label{fig:vel_tests}
\end{figure*}

The velocity radial profile that reproduces best our observations is represented by a static central core (v=0 at r$<$ 3000 au) and an infalling outer gas layer (v=-0.5 km s$^{-1}$ at r$>$ 3000 au).
Our tests using different velocity radial profiles are shown in Figure~\ref{fig:vel_tests}. The asymmetry in the N$_2$H$^+$ 3-2 line cannot be reproduced with a static core (green curve in Figure~\ref{fig:vel_tests}). 
Most of the lines, but particularly the high-$J$ transitions of both N$_2$H$^+$ and N$_2$D$^+$, cannot be reproduced if we assume a static envelope and an infalling central part of the core (v=-0.5 km s$^{-1}$ at r$<$ 3000 au, see the blue dashed curve in Figure~\ref{fig:vel_tests}.\\

\begin{figure*}

\begin{center}
\includegraphics[width=18.5cm]{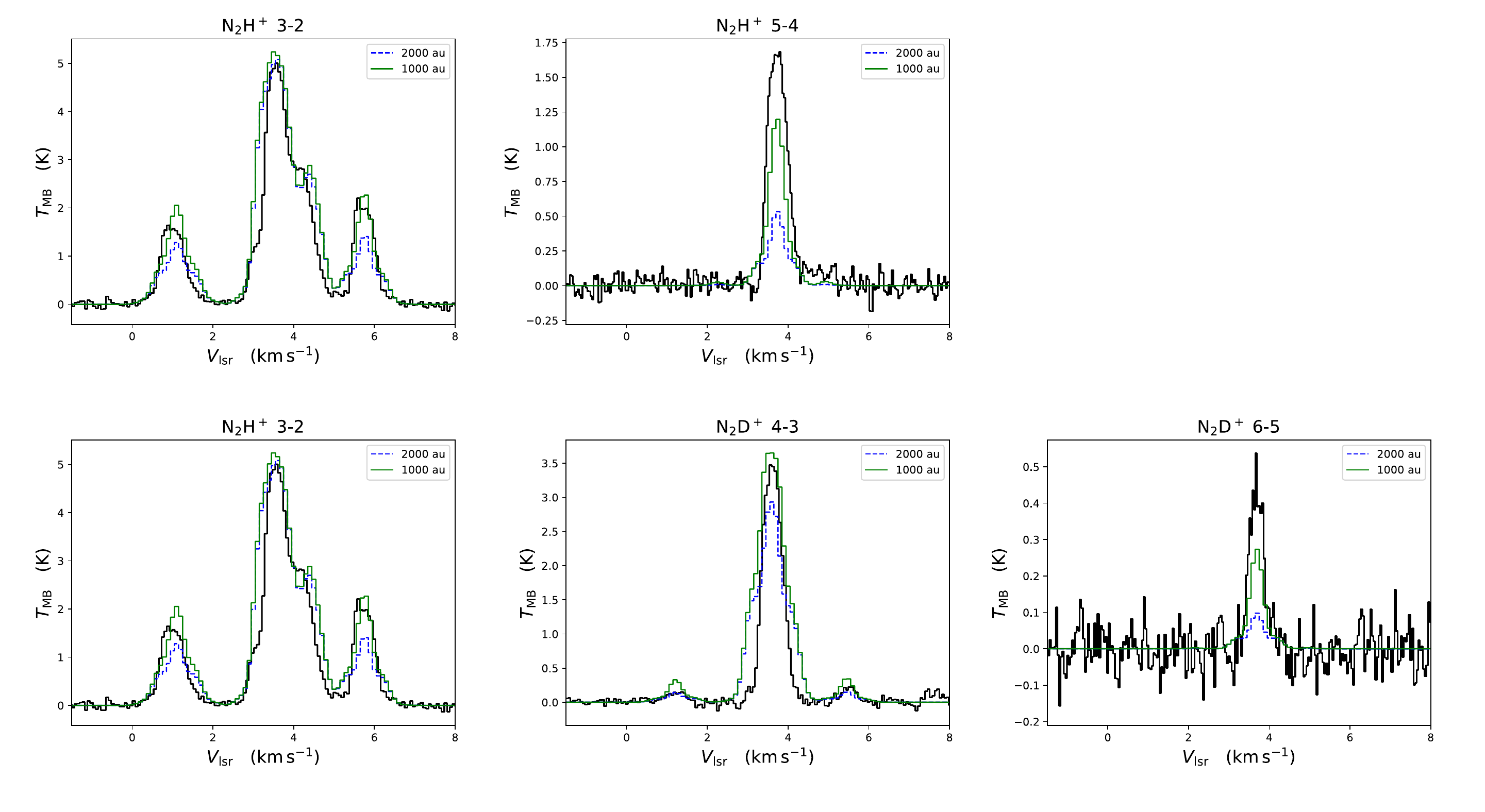}
\end{center}
\caption{Results of tests on the effects of depletion. The blue dashed curve shows the results of the non-LTE modelling done assuming a complete depletion in the inner 2000 au of the core, while the green curve assumes a complete depletion in the inner 1000 au of the core. }
\label{fig:depletion_tests}
\end{figure*}

The abundance profile that reproduces best our observations is a step function with values of 4.5$\times$10$^{-10}$ for N$_2$H$^+$ and 2$\times$10$^{-10}$ for N$_2$D$^+$ in the inner 3000 au and a drop of 30\% of the central values at radii larger than 3000 au. It is however expected for molecules to freeze-out in the central regions of pre-stellar cores because of the high densities and low temperatures \citep{caselli22}. 
Our tests assuming a complete depletion in the central 2000 and 1000 au are shown in Figure~\ref{fig:depletion_tests}. The high-$J$ transitions of N$_2$H$^+$ and N$_2$D$^+$ are poorly reproduced if assuming a complete depletion in the central 2000 au (dashed blue curve in Figure~\ref{fig:depletion_tests}). It is important to note that these transitions have a high critical density ($\sim$10$^7$ cm$^{-3}$) and they are presumably emitting only in the dense gas at the centre of the pre-stellar core. A significant improvement can be seen if we assume a smaller radius for the complete depletion, 1000 au (green curve in Figure~\ref{fig:depletion_tests}), although the high-$J$ transitions are still better reproduced without depletion (red curve in Figure~\ref{fig:LOC}). 
However, given the angular resolution of our $Herschel$ observations ($\sim$2700 au), we would need a better resolved physical structure to make a more conclusive statement on the complete freeze-out radius for N$_2$H$^+$ and N$_2$D$^+$ in I16293E.\\

\begin{figure*}

\begin{center}
\includegraphics[width=18.5cm]{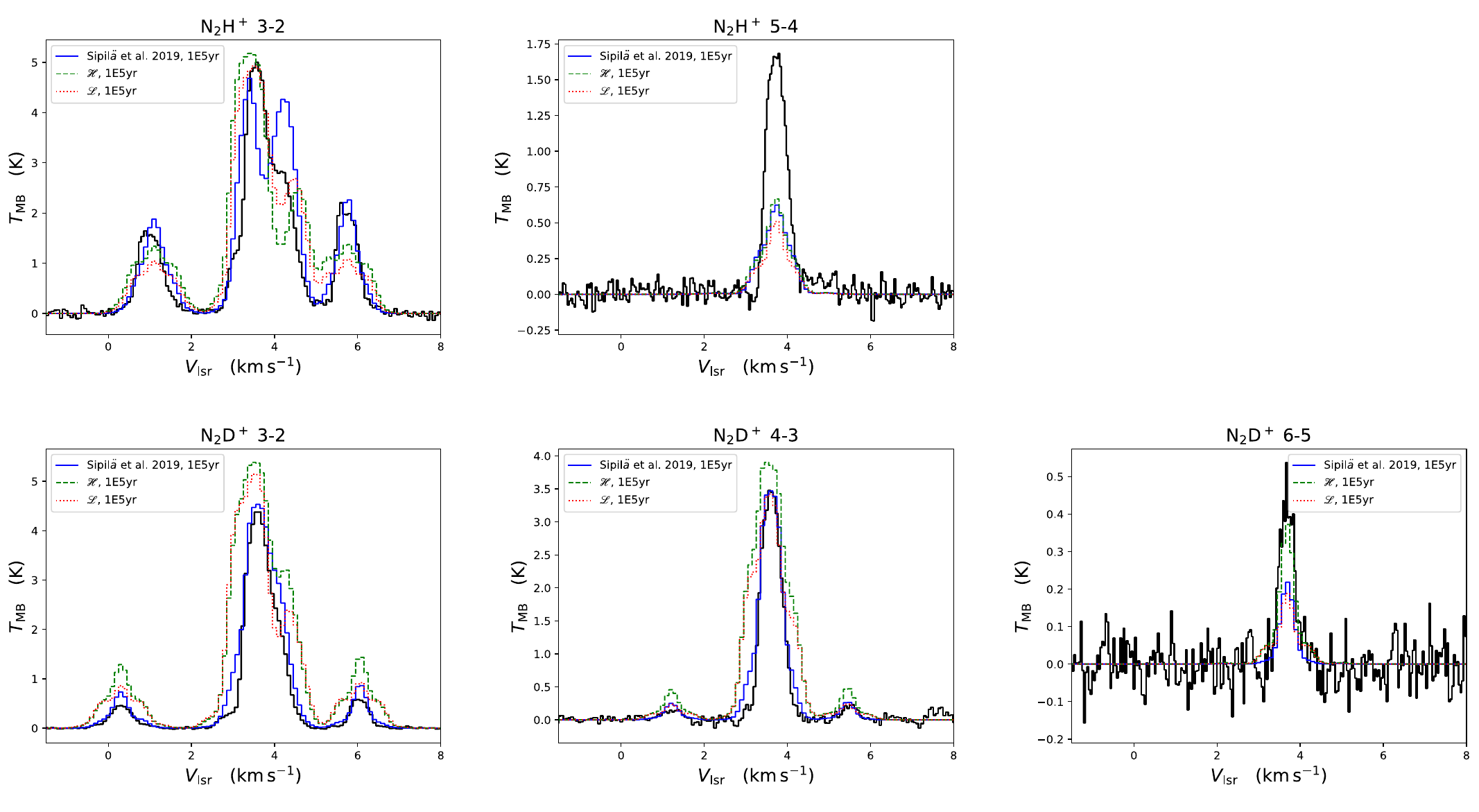}
\end{center}
\caption{Results of tests using abundance profiles from chemical models. The blue curve shows the results of the non-LTE modelling using the abundance profiles of N$_2$H$^+$ and N$_2$D$^+$ predicted from the chemical model presented in \cite{sipila19}. The green dashed and red dotted curves instead use a chemical model with a radial dependence of the cosmic ray ionisation rate,  with the high model ($\mathscr{H}$) and with the low model ($\mathscr{L}$) from \cite{padovani18}, respectively. }
\label{fig:olli}
\end{figure*}

\begin{figure*}

\begin{center}
\includegraphics[width=15cm]{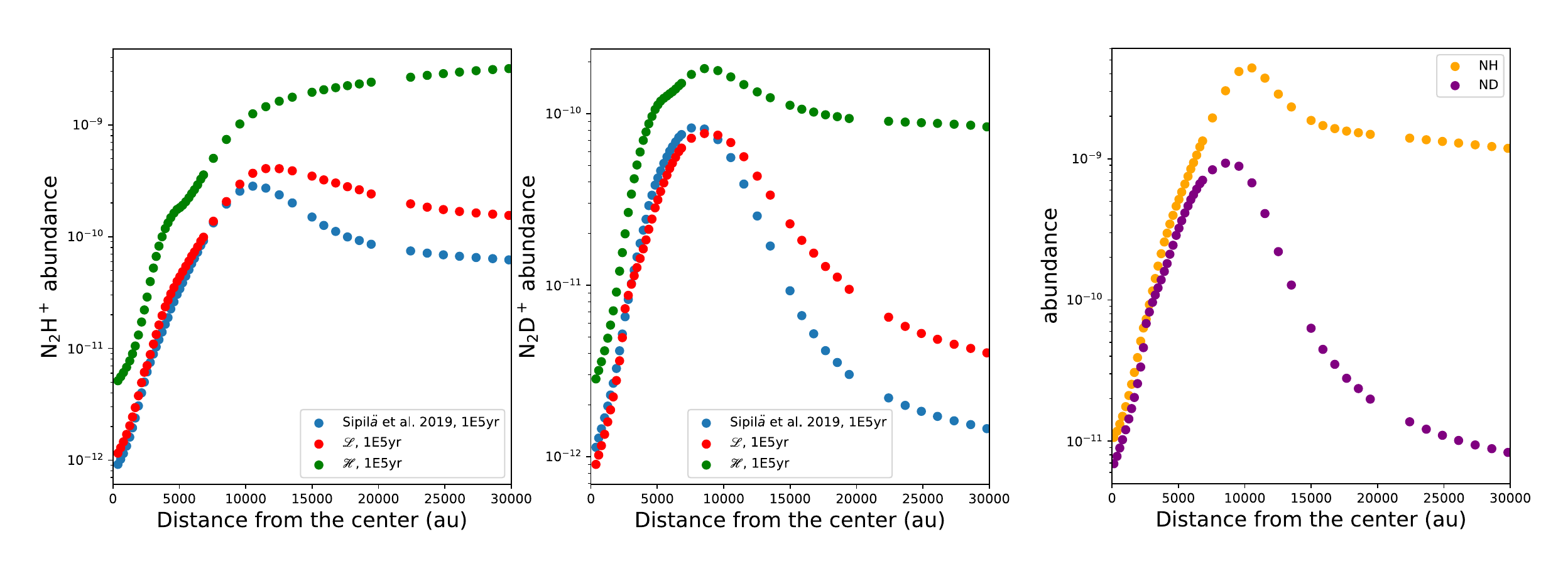}
\end{center}
\caption{Left and central panels: abundance profiles from different models for N$_2$H$^+$ and N$_2$D$^+$. Right panel: abundance profiles for ND and NH from \cite{sipila19} at 10$^5$yr.}
\label{fig:profiles}
\end{figure*}

Figure~\ref{fig:olli} shows the results of radiative transfer models using the radial  abundance profiles from the chemical models described in \cite{sipila19}. We furthermore tested the effects of a radial dependence of $\zeta$ following the approach described in \cite{padovani18} and used in \cite{redaelli21}.
The abundance profiles that have been used are shown in Figure~\ref{fig:profiles}.

\section{Dust temperature and $\htwo$ column density maps of I16293E}

\begin{figure*}
\centering
\unitlength=1mm
\begin{picture}(160,70)(0,0)
\put(-10,-3){
\begin{picture}(0,0) 
\includegraphics[width=9cm,angle=0]{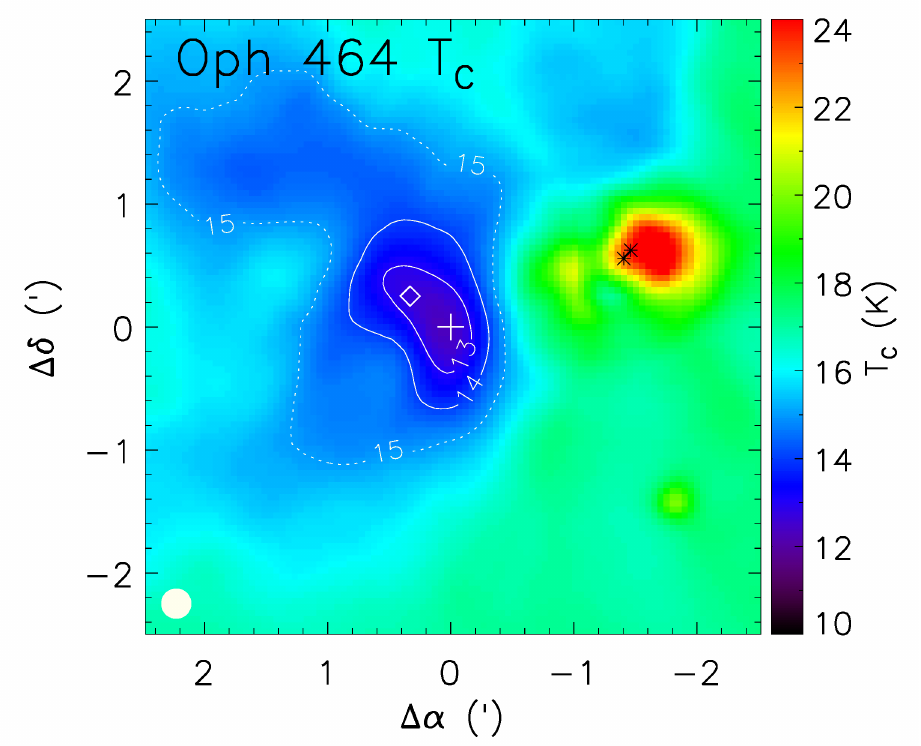}
\end{picture}}
\put(80,0){
\begin{picture}(0,0) 
\includegraphics[width=9cm,angle=0]{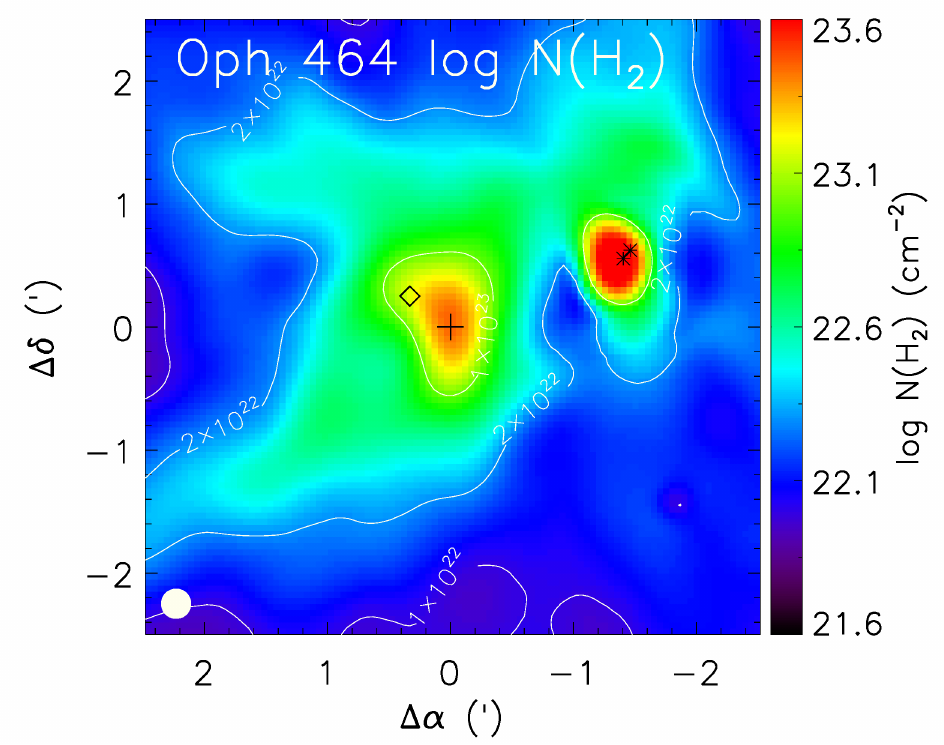}
\end{picture}}
\end{picture}  
\caption[]{Dust colour temperature ($T_{\rm C}$) and $\htwo$ column density ($N(\htwo)$) maps of the IRAS 16293 region derived using {\sl Herschel} PACS images at 70, 100, and $160\,\micron$, and deconvolved SPIRE HiRes images at 250, 350, and $500\,\micron$. The position observed in the present work is indicated with a plus sign, and the $\dtwohplus$ peak discovered by \cite{pagani24} is indicated with a diamond. The protostellar core IRAS 16293-2422 dominates the right side of both images. The positions of the protostars IRAS16293-2422 A and B are shown with asterisks. The column density at the (0,0) is $2.8\times10^{23}\persqcm$ and the $T_{\rm C}$ there is 12.3\,K. The effective beam of the SPIRE HiRes map at $500\,\micron$ is indicated in the bottom left corner of the image.}
\label{tc_and_col_maps}
\end{figure*}

The dust colour temperature ($T_{\rm C}$)  and $\htwo$ column density ($N(\htwo)$) maps of the I16293E core and the adjacent protostellar core IRAS 16293-2422 are shown in Fig.~\ref{tc_and_col_maps}. These are derived from {\sl Herschel}/PACS  and SPIRE/HiRes images at the wavelengths $\lambda=70$, 100, 160, 250, 350, and $500\,\micron$ by fitting a grey body function to aligned surface brightness maps.  Colour corrections recommended in {\sl Herschel} documentation\footnote{https://www.cosmos.esa.int/web/herschel/legacy-documentation} were applied to the observed surface brightness values in the iterative solution, needed to take into account the dependence of the effective bandwidth of the instrument on the temperature of the source. We assumed that the dust opacity at $250\,\micron$ is $\kappa_{250\,\micron} = 0.1\,{\rm cm^2\,g^{-1}}$, and that the dust emissivity index is $\beta=2$. This assumption approximately corresponds the dust opacity model of \cite{ossenkopf1994} for unprocessed dust particles with thin ice coatings. The positions of the dust emission peak in I16293E, the $\dtwohplus$ peak discovered by \cite{pagani24}, and the protostars IRAS 16293-2422 A and B are indicated in the maps.

\end{appendix}

\end{document}